\documentstyle[prl,aps,twocolumn]{revtex}
\input{epsf}
\begin{document}
\draft
\title{Induced spin polarisation in a ferromagnetic
gadolinium-yttrium alloy}

\author{J.A. Duffy$^{1}$, S.B. Dugdale$^{2,3}$, J.E. McCarthy$^{4}$,
M.A. Alam$^{3}$, M.J. Cooper$^{1}$, S.B. Palmer$^{1}$ and T.
Jarlborg$^{2}$.}

\address{$^{1}$ Department of Physics, University of Warwick,
Coventry, CV4 7AL, UK}

\address{$^{2}$ D\'epartement de Physique de la
Mati\`ere Condens\'ee, Universit\'e  de Gen\`eve, 24 quai Ernest Ansermet,
CH-1211 Gen\`eve 4, Switzerland}

\address{$^{3}$ H.H. Wills Physics Laboratory, University of Bristol,
Tyndall Avenue, Bristol BS8 1TL, UK}

\address{$^{4}$ ESRF, BP220, F-38043 Grenoble Cedex, France}

\date{\today}
\maketitle
\begin{abstract}
The first direct evidence of an induced spin moment in
Gd$_{62.4}$Y$_{37.6}$ is presented.  This additional moment, of $0.16 \pm
0.03 \mu _{B}$, arises from polarisation of Y electrons in the alloy.  The
moment was detected in a Compton scattering experiment via the measurement
of the one dimensional projection of the momentum space electron spin
density in Gd and in the alloy.  The result is consistent with theoretical
predictions calculated using the LMTO method within the local spin density
approximation.
\end{abstract}


Oscillatory exchange coupling \cite{majkrzak:86} was first observed in
superstructures of Gd and Y. This was quickly followed by
observations of giant magnetoresistance in other similar multilayer
systems. The coupling between successive ferromagnetic Gd layers is
thought to rely upon the spin-polarisation of the Y layers
\cite{chappert:91}, although this polarisation has never been observed.
Given that yttrium plays a similar important role in bulk Gd--Y alloys, 
we have studied the latter in order to such for evidence of such an
induced moment.  The bulk alloy system
exhibits interesting magnetic behavior, having three ordered phases
\cite{palmer:85}, and is an excellent system for studying the
magnetic interactions.  At high Gd concentrations (above 70\% Gd) the
alloy has different ferromagnetic phases at low and high
temperature. In compositions containing less than 60\% Gd the alloy 
exhibits a helical antiferromagnetic phase.  In the 
intermediate compositions there is a delicate balance between the three
phases. For a large composition range, the total magnetic moment is
greater than would be expected simply from the dilution of the Gd
\cite{thoburn:58}.  There has been considerable effort made towards an
understanding of the magnetic structures of the Gd-Y alloys, but the
nature of the excess moment has not hitherto been fully resolved.
Moreover, the
physics of the magnetic ordering of the Gd-Y alloy system is also
considered to arise from an indirect exchange interaction involving the Y
electrons. In this Letter we present the first experimental evidence for
the polarisation of the Y electrons, in the form of an excess {\it spin}
moment in
ferromagnetic Gd$_{62.4}$Y$_{37.6}$.

Gadolinium metal is a 4$f$ ferromagnet, with a magnetic moment of
$7.63\mu_{B}$ \cite{roeland:75} and a Curie temperature of $294K$.  The
moment comprises $7\mu_{B}$ from the half-filled 4$f$ shell, plus an
induced conduction electron moment of $0.63\mu_{B}$.  As predicted by
Hund's rules, there is no orbital component $(L=0)$. Gadolinium has the
hcp structure, and the moment aligns along the c-axis down to $235K$,
below which it becomes canted \cite{coqblin:77}. Yttrium has the same
structure and a similar atomic volume to Gd, and hence the alloys readily
form a continuous solid solution with only small changes in their lattice
parameters. The non-magnetic Y ``impurities'' might be thought to have
very little effect on the magnetic properties, acting simply as a diluent.
However, Thoburn {\it et al.} \cite{thoburn:58} showed that the addition
of Y does not simply monotonically reduce the
total moment; an extra contribution is apparent in the ferromagnetic
alloy. The behaviour of the total moment could be explained in two ways:
either the presence of yttrium modifies the crystal field, resulting in an
{\it orbital} contribution to the moment, or the hybridised
conduction bands in the alloy enable a larger {\it spin} moment to
be induced.

The 4$f$ electrons of Gd are highly localised, and the magnetic ordering in
both the Gd metal and in the Gd-Y alloy arises from an indirect exchange
interaction mediated
via the conduction electrons.  This RKKY-type interaction is explained in
terms of the wave-vector dependent susceptibility, $\chi({\bf q})$
\cite{kasuya:66}.  In pure Gd, $\chi({\bf q})$ has a maximum at ${\bf
q}=0$, leading to the observed ferromagnetic ordering.  If this maximum is
at a non-zero ${\bf q}$, because of yttrium induced \cite{dugdale:97}
Fermi surface nesting in the alloys \cite{fretwell:99}, then a more
complex arrangement of spins may form \cite{palmer:85,freeman:72}.  The
RKKY-type interaction relies on the polarisability of the Y conduction
bands.  However, it has recently been proposed that the
additional moment can be accounted for purely by considering orbital
contributions originating from the modified crystal field
\cite{foldeaki:95}.  This was reasoned from the experimental behaviour of
the effective Land\'e $g$ factor and total angular momentum, $J$, which is
compatible with the assumption that the crystal field is changed by the
presence of Y, permitting spin-orbit coupling to induce the orbital moment.
The presence of a spin moment in the low temperature ferromagnetic
(``ferro II'') phase would indicate that the exchange-splitting persists
and that the Y bands are polarised, obviating the need for a large orbital
contribution in the alloy.  The goal of the experiment reported here was
to determine whether there is indeed an extra spin moment contribution to
the magnetisation in the low temperature ferro II phase.  

The experiment was performed using magnetic Compton scattering, a uniquely
sensitive probe of the spin component of the magnetisation.  The Compton
profile, $J(p{_z})$, is defined as the 1-dimensional projection of the
electron momentum distribution, $n({\bf p})$, 

\begin{equation}
J(p {_z} ) = \int \int n({\bf p}) 
 {\rm d}p_{x} {\rm d} p_{y}.
\label{ccp}
\end{equation}

and the integral of $J(p{_z})$ is  simply the total number of electrons 
per unit cell.  The profile can be obtained experimentally from the
energy spectrum of the inelastically scattered photons.  This is achieved
by exploiting the Compton effect,
in which monochromatic photons scattered through a
given angle by stationary electrons would have a single energy determined
purely
by the scattering angle.  However, because bound electrons must have some
distribution of momenta, the photon energy is Doppler broadened into an
energy distribution. This is related to the Compton profile, defined
above, via the scattering cross-section\cite{holm:88}, within the
impulse approximation \cite{platzman:70}.  If
the photons impinging on a sample have a component of circular
polarisation, then a small spin dependence appears in the scattering cross
section\cite{bell:96}.  Reversing either the photon polarisation or the
magnetisation of the sample changes the sign of the spin-dependent signal,
which enables the spin part to be isolated.  The resultant profile,
known as the magnetic Compton profile (MCP), is a projection of the
momentum density of only those electrons with unpaired spins,

\begin{equation}
J_{\mbox{mag}}(p {_z} ) = \int \int \left(n\!\!\uparrow\!\!({\bf p}) -
n\!\!\downarrow\!\!({\bf p}) \right)
 {\rm d}p_{x} {\rm d} p_{y}.
\label{mcp}
\end{equation}

Here, $n\!\!\uparrow\!\!({\bf p})$ and $n\!\!\downarrow\!\!({\bf p})$
are the momentum dependent spin densities.  The area under the MCP 
is equal to the number of unpaired electrons, that is, the 
total spin moment per formula unit:

\begin{equation}
\int_{-\infty}^{\infty} J_{\mbox{mag}}(p_{_z} ) {\rm d} p{_z} =
\mu_{\mbox{spin}}.
\label{intmcp}
\end{equation}

Magnetic Compton scattering is now an established technique for probing
momentum space spin densities and band structures in magnetic materials
\cite{sakai:96,cooper:97}.  Within the impulse approximation, the method
is solely sensitive to $\it{spin}$ magnetic moments
\cite{sakai:96,cooper:92,lovesey:96}; that is to say, the orbital moment
is not measured \cite{carra:96}.  The value of magnetic Compton scattering
lies in its uniform sensitivity to the whole of the spin-resolved electron
momentum distribution. Since the MCP is a
difference between Compton profiles, the contributions from the
non--magnetic electrons and from unwanted systematic sources disappear.
Spin-polarised positron angular correlation
experiments also probe the spin density \cite{berko:68,manuel:91}, but are
subject to both positron--electron correlation effects and repulsion of
the positron by the positive ion cores, so that the positron does not
sample electrons in all states equally \cite{west:95}.  
Furthermore, the incoherent nature of the Compton scattering process means
that the electron density distribution can be sampled at all momenta,
notably at the low momenta where the conduction electrons contribute.

The [0001] MCP for Gd$_{62.4}$Y$_{37.6}$ was measured on the high energy
x-ray beamline at the ESRF. The experiment was performed in
reflection geometry \cite{duffy:98} with an incident beam energy of 200keV,
selected by the $\{ 311 \}$ reflection of a Si monochromator, and a
scattering angle of 168$^{\circ}$.  The samples were 5mm diameter
$\times$ 1.3mm thick disks and were oriented so that the resolved
direction was within $\pm 2^{\circ}$ of [0001].  The temperature of the
samples was maintained at $70 \pm 2$K.  At present, it is difficult to
reverse the polarisation of
the synchrotron x-ray beam, but in soft ferromagnets like these, the
sample's magnetisation can be easily reversed.  Here, the magnetisation was
kept alternately parallel and antiparallel to [0001] with a 0.96T rotating
permanent magnet.  The energy spectrum of the scattered x-rays was
measured by a solid-state Ge detector. The momentum resolution obtained was
0.44 atomic units (a.u., where 1 a.u. = 1.99$\times$10$^{-24}$ kg m
s$^{-2}$).  The total number of counts in each of the charge profiles was
1.5 $\times 10^{8}$, resulting in 3.7 $\times 10^{6}$ in the MCP with a
statistical precision of $\pm 3\%$ at the magnetic Compton peak in a bin of
width 0.09 a.u.  The usual correction procedures for the energy
dependence of the detector efficiency, for absorption, relativistic
scattering cross-section and magnetic multiple-scattering
were applied
and after checking that the profiles were symmetric about zero momentum,
the MCPs were folded about this point to increase the effective statistical
precision of the data.

\begin{figure}
\epsfxsize=230pt
\epsffile{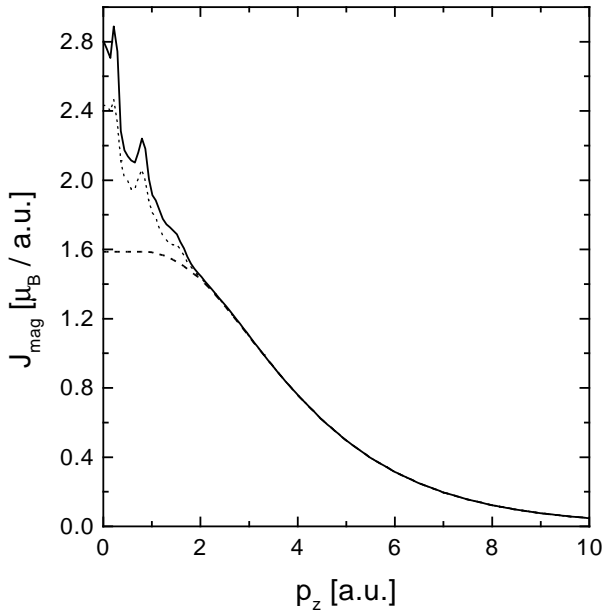}
\caption{Calculations of the magnetic Compton profile of Gd (dots)and
Gd$_{62.5}$Y$_{37.5}$ (solid line) resolved along [0001], performed using a
sixteen atom supercell.  Also shown is the equivalent free atom profile for
Gd 4$f$ electrons (dashes).}
\label{theory}
\end{figure}

The experiments were complemented by LMTO band-structure calculations
\cite{andersen:75} performed within the local spin density approximation (LSDA)
\cite{gunn:76}.  The authors recently demonstrated the ability of this
technique to predict the magnetic Compton profiles of ferromagnetic Gd
metal and the details of the calculations are described
more fully there \cite{duffy:98}.  In order to predict the MCP of the
disordered Gd$_{62.4}$Y$_{37.6}$ alloy, a sixteen atom supercell approach was
adopted.  By taking ten Gd and six Y atoms, a effective composition of
Gd$_{62.5}$Y$_{37.5}$ was obtained, very close to that of the
measured sample.  The results were essentially unchanged for
different configurations of the sixteen atoms.  In order to provide a
suitable comparison, the pure Gd MCP was also calculated in the supercell
and it was found to be essentially identical to that from the standard
calculation.  The lattice parameter and $c/a$ ratio were 6.871 a.u. and 1.59
for Gd, and 6.8758 a.u. and 1.584 for the alloy.  The
predicted moments and their associated characters for Gd and
Gd$_{62.5}$Y$_{37.5}$ are presented in Table \ref{bandspin}.  It can be
seen that an extra spin moment of $0.35 \mu_{B}$ is expected per Y atom,
mainly on the $p$- and $d$-like electrons, and this corresponds to $0.14
\mu_{B}$ per cell.  The predicted [0001] MCPs are presented in Figure
\ref{theory}, where the 4$f$ moments have been normalised to account for
the Gd dilution in the alloy.  This clearly shows that the conduction
electron contribution increases relative to the 4$f$ moment in the alloy.

\begin{figure}
\epsfxsize=230pt
\epsffile{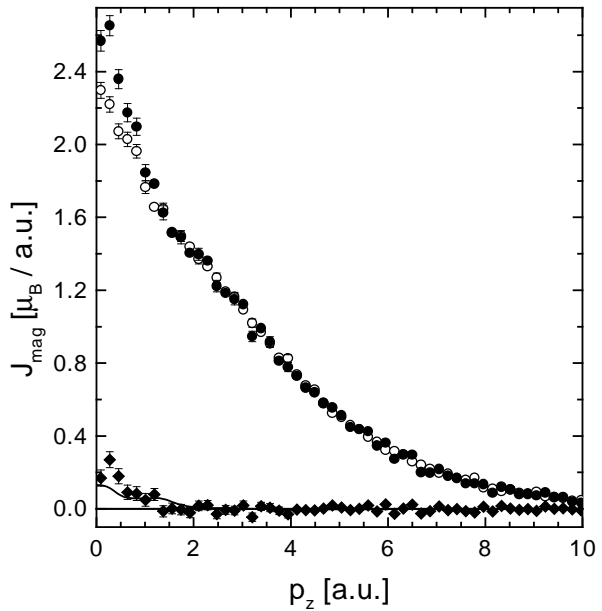}
\caption{The experimental magnetic Compton profiles of Gd, normalised to
$7.62 \mu_{B}$ (open circles) and Gd$_{62.4}$Y$_{37.6}$ (solid circles),
scaled to $4f$ tails of the Gd profile.  The diamonds represent the
difference in Bohr magnetons per a.u. for a formula unit of the alloy. The
difference in the theoretical profiles (Figure 2),
convoluted with a Gaussian with FWHM = 0.44 a.u. to represent the
experimental resolution, is presented as a solid line.}
\label{experiment}
\end{figure}

In Figure \ref{experiment}, we present the experimental MCPs for Gd and
Gd$_{62.4}$Y$_{37.6}$, normalised according to their 4$f$ moments.  These
are presented together with the theoretical predictions after convolution
with a Gaussian of FWHM 0.44a.u. to simulate the experimental resolution.
The Gd profile is analysed in more detail in \cite{duffy:98}.  The salient
features here are the narrow conduction electron contributions at low
momentum superimposed on the broader 4$f$ profile, as expected.
Comparison of the experimental profiles for the pure metal and the alloy
shows that there is a small, but genuine difference for $p_{z}<2$ a.u.,
consistent with that predicted by theory. 
By integrating over $p_{z}$, the value of the induced moment can be
calculated as,

\begin{equation}
\Delta\mu_{spin} = 0.16 \pm 0.03 \mu_{B}.
\label{result}
\end{equation}

In order to investigate the robustness of this small difference, we
performed a number of checks during the data analysis.  The difference is
present in the raw, uncorrected data.  The corrections applied to the Gd
and Gd$_{62.4}$Y$_{37.6}$ data sets are essentially the same, since both
measurements were performed with the same setup, and the detector
efficiency and scattering cross-sections are identical in both cases.  The
multiple scattering contribution was calculated to be an order of magnitude
smaller than the measured difference signal.  The absorption correction
was calculated to simulate possible misalignments in the setup, but the
results remained unchanged.  It should be noted that near $p_{z}=0 a.u.$
this correction is almost a linear function of $p_{z}$ and any error will
be unlikely to affect the difference observed, especially when the data
are folded about the origin.  Hence, the result is robust to any
reasonable variations in the corrections applied.
 
In conclusion, we have presented the first direct evidence of an additional
spin moment in ferromagnetic Gd$_{62.4}$Y$_{37.6}$ alloy, corresponding to
a polarisation of the Y band electrons.  It is unaffected by the
corrections that need to be applied to the experimental data.
Band-structure calculations performed within the LSDA are consistent with
our result.  It should be remembered that the presence of this spin moment
does not rule out the existence of an additional small orbital moment.
However, in contrast with the interpretation of Foldeaki {\it et al.}
\cite{foldeaki:95}, we conclude that, irrespective of any orbital
contribution, there is indeed a substantial polarisation-induced spin
moment of 0.1$6\pm 0.03 \mu_{B}$ in the alloy.

We would like to thank the ESRF for allocation of beam time, and the EPSRC
(UK) for generous financial support.  One of us (SBD) would like to thank
the Royal Society (UK) and the Swiss National Science Foundation for a
Fellowship.  We also thank Martin Lees for performing magnetisation
measurements at Warwick.

%
%

%
%

\begin{table}
\caption{Calculated partial spins in Bohr magnetons per atom for pure Gd
and for  Gd and Y in Gd$_{62.5}$Y$_{37.5}$.  Also shown are the moments
per formula unit of the alloy.}
\label{bandspin}  
\begin{tabular}{cccccc}
& $s$ & $p$ & $d$ & $f$ & Total \\
Gd (pure)  & 0.025 & 0.141 & 0.580 & 6.94 & 7.64 \\
Gd (alloy)   & 0.032 & 0.153 & 0.525 & 6.85 & 7.56 \\
Y (alloy) & 0.000 & 0.124 & 0.205 & 0.022 & 0.350 \\
Total     & 0.020 & 0.142 & 0.405 & 4.29 & 4.91 \\
\end{tabular}
\end{table}

\end{document}